\documentclass[12pt,a4wide]{article}
\usepackage{graphicx}

\begin{document}

\newcommand{\lsim}{\mbox{\raisebox{-.9ex}{~$\stackrel{\mbox{$<$}}{\sim}$~}}}
\newcommand{\gsim}{\mbox{\raisebox{-.9ex}{~$\stackrel{\mbox{$>$}}{\sim}$~}}}

\begin{center}
{\Large\bf Steep Eternal Inflation and the Swampland}
\end{center}

\medskip

\noindent
\begin{center}
{\large Konstantinos Dimopoulos$^*$}
\end{center}

\medskip

\begin{center}
\noindent
$^*${\small\em Consortium for Fundamental Physics, Physics Department,}\\
{\small\em Lancaster University, Lancaster LA1 4YB, UK}\\
{\small e-mail: {\tt k.dimopoulos1@lancaster.ac.uk},}\\
\end{center}

\medskip

\begin{abstract}
I investigate whether eternal inflation is possible when the inflaton
originally travels upslope the scalar potential and inevitably reaches a 
turn-around point where its classical kinetic energy momentarily vanishes.
This behaviour occurs regardless of the steepness of the potential slope. Such 
steep eternal inflation, if achieved, would satisfy the recent swampland 
conjecture and allow eternal inflation in the string landscape.
\end{abstract}

Eternal inflation is a mode of inflation where the variation of the inflaton 
field is not determined by its classical motion in field space, but by its 
quantum fluctuations \cite{eternal,eternallit}. 
Typically, this corresponds to regions of the scalar 
potential, where the potential slope is shallow and/or the Hubble friction is 
huge. These regions are called ``diffusion zones'' because, under the influence 
of the quantum fluctuations, the inflaton random walks inside the diffusion zone
(i.e. its expectation value diffuses in space), assuming equiprobable values 
throughout the diffusion zone, oblivious of the scalar potential and its slope. 
Hence, at some locations in space, it is possible for the inflaton to never 
exit the diffusion zone, in which case inflation never ends. At other locations 
however, the inflaton exits the diffusion zone and ceases to be dominated by 
its quantum fluctuations. In the latter case, the inflaton classically rolls 
down the scalar potential until inflation ends. In this picture, the global 
universe is undergoing eternal inflation, with the inflaton in the diffusion 
zone. However, occasionally, in some locations, the inflaton exits the 
diffusion zone by chance (the quantum fluctuations are random) and classical 
inflation continues, which eventually ends. Thus, in these locations, we have 
the formation of a so-called pocket universe, where inflation ends and 
subsequent evolution ensues. Our Universe is one such pocket universe. So, 
amids eternally inflating space, pocket universes are continuously spawned, 
with our Universe being one of them.

The original eternal inflation proposal considered a chaotic monomial potential
\cite{eternal},
where eternal inflation occurs up the slope near the Planck density, when 
Hubble friction is huge. In this potential there is only one minimum so all the
pocket universes would be identical and would correspond to the locations where 
the diffused inflaton finds itself at low values and exits the diffusion zone.
Soon it was realised that the situation would differ with more complicated 
potentials, featuring multiple minima. In such a potential the fate of the 
inflaton, when exiting the diffusion zone, is not unique. Indeed, after eternal 
inflation the inflaton may find itself in a different minimum at a different 
location in space. In this case the pocket universes are not identical, but 
belong to classes corresponding to different minima. This possibility is 
maximised in the string landscape \cite{landscape}, 
which features a huge multiple of minima,
in the form of valleys and sinks in the multi-field scalar potential of the 
moduli space \cite{landscape+}. 
Hence, it has been theorised that most, if not all the minima of 
the string landscape are populated through eternal inflation, with pocket 
universes giving rise to many different universes, one of which is our own. 
This is called the string multiverse. The multiverse idea allowed authors to 
use anthropic arguments to make sense of a number of tunings apparent in our 
Universe, that are unexplained but necessary for life to 
exist~\cite{landscape}.

Recently, however, a conjecture was put forward, which requires the slope of the
scalar potential in string theory to be bounded from below \cite{swampconj}. 
In rough terms, it is conjectured that the slope of the potential cannot be 
shallow, especially on ``high ground'' where the potential density is large as 
is the case of primordial inflation. Effective field theories which violate this
conjecture cannot be resulting from string theory, so they are not part of the 
string landscape but instead of the so-called swampland.%
\footnote{For cosmological consequences of this conjecture see 
Ref.~\cite{swampcosmo,linde}.
For critical discussion on this conjecture see Ref.~\cite{linde,swampcrit}.
For discussing Higgs physics in terms of this conjecture see
Ref.~\cite{Higgsswamp}.}
This conjecture may undermine 
eternal inflation, which seems to require a shallow potential slope. Indeed, the
issue was studied in Ref.~\cite{swampeternal} and it was 
found that the proposed conjecture does not allow eternal inflation in string 
theory. As a result, the string landscape cannot be populated and the formation 
of the string multiverse is questionable. 

In this letter, however, we investigate a different mechanism which may lead to 
eternal inflation and could operate with steeper potential. The idea is that 
if the initial conditions of the inflaton send it upwards in the scalar
potential, then inevitably, its climb stops at some point, where it turns around
and rolls downwards. At the turn-around moment, its classical kinetic density is
zero, which means that its quantum fluctuations, no matter how small, may take 
over. If this were so, then there would be locations in space where the quantum 
fluctuations of the inflaton conspire to keep it up the scalar potential 
forever. Can this lead to eternal inflation with a steep potential? If it did, 
then the formation of the string multiverse would not be threatened by the 
recent swampland conjecture.

We work in Einstein gravity only. In the following, we use natural units, where 
$c=\hbar=1$ and Newton's gravitational constant is \mbox{$8\pi G=m_P^{-2}$}, 
with \mbox{$m_P=2.43\times 10^{18}\,$GeV} being the reduced Planck mass. 

First, let us briefly review standard eternal inflation. The idea is as follows.
During inflation, gravitational particle production generates perturbations
of all light, non-conformally invariant fields. In the case of a light, 
minimally coupled scalar field, its perturbations are given by the Hawking 
temperature in de Sitter space \mbox{$\delta\phi=H/2\pi$}, which are generated 
per Hubble time $\delta t=H^{-1}$, where $H$ is the Hubble parameter 
(approximately constant in inflation). Thus, due to these quantum fluctuations 
the scalar field is constantly displaced during inflation. This displacement is 
negligible if it is overwhelmed by the classical variation of the field, as it 
rolls down the scalar potential. However, if the kinetic density of the 
classical variation is very small, then the quantum displacement takes over and 
determines the field evolution. Because quantum fluctuations are random and 
oblivious of the scalar potential, it is equally
possible to push the field towards larger potential values as well as smaller 
potential values. Thus, if the quantum variation of the field dominates its 
classical variation and if this field drives the inflationary expansion (the 
inflaton field) then a fraction of the inflating space will continue 
inflating forever as the field is kept at high potential values. 

The criterion 
is obtained by comparing the classical to the quantum variation of the field
\begin{equation}
\frac{\delta\phi}{\delta t}>|\dot\phi|
\;\Rightarrow\;|V'|<\frac{3}{2\pi}H^3
\label{H3}
\end{equation}
where we also assumed the (classical) slow-roll equation 
\mbox{$3H\dot\phi\simeq -V'$}, where $V$ is the scalar potential density
with the dot denoting time derivative and the 
prime denoting derivative with respect to the scalar field. The above suggests 
that the quantum fluctuations may lead to eternal inflation if the inflaton 
finds itself at a part of the potential where the potential slope is small 
enough to satisfy the bound in Eq.~(\ref{H3}).

Recently, however, a suggestion was put forward that the effective field 
theories coming down from string theory (so part of the string landscape) cannot
feature a scalar potential with too shallow a slope. Instead it was conjectured
that there is a bound on the scalar potential slope of the form
\begin{equation}
\frac{|V'|}{V}>\frac{c}{m_P}\,,
\label{swampland}
\end{equation}
where $c$ is a positive constant {\em of order unity}. Using the Friedmann 
equation in slow-roll \mbox{$V\simeq 3H^2m_P^2$} and combining Eqs.~(\ref{H3}) 
and (\ref{swampland}) it is straightforward to find that one may feature eternal
inflation only when \cite{swampeternal}
\begin{equation}
c<\frac{H}{2\pi m_P}\ll 1\,,
\label{crange}
\end{equation}
where we consider \mbox{$H<m_P$} to avoid quantum gravity corrections.

Because we see that \mbox{$c\ll 1$} is needed, it seems that eternal inflation 
is not possible in the string landscape. Instead, effective field theories which
allow eternal inflation find themselves in the swampland.
As a result, eternal inflation cannot be used to populate the 
string landscape and lead to the formation of the string multiverse.
Employing the multiverse hypothesis enabled 
explaining away the cosmological constant problem through anthropic arguments.
Without eternal inflation, this seems not as possible.

Can the above situation be remedied? We discuss here a different possibility
to realise eternal inflation, even without the requirement of a shallow enough
slope in the scalar potential. We call this possibility Steep Eternal Inflation,
which if it works, may allow the realisation of the string multiverse without
violating the conjecture in Eq.~(\ref{swampland}). 

The basic idea is simple. 
Suppose that we are in a part of the potential with a steep slope. If the 
initial conditions are such that the kinetic density of the scalar field is 
sending the field upwards on the slope,\footnote{There are string theory 
setups where scalar fields may be forced to climb up the potential after the 
Big Bang \cite{upclimb}.} then the climbing field reaches a 
maximum and turns-around rolling down the slope. At the turn-around, the 
classical kinetic density of the field becomes momentarily zero. Thus, it is 
likely that near the turn-around the variation of the field becomes dominated
by its quantum fluctuations.
We investigate whether this can lead to eternal inflation, where there are 
always parts of the Universe where the quantum kicks conspire to keep the field
up the slope. Given that, after the exit from spacetime foam, chaotic initial 
conditions may well lead to the initial kinetic density sending the field 
upslope, we believe that the the above scenario can be naturally realised.

To model the evolution near the turn-around we Taylor expand the scalar 
potential near the turn-around point as
\begin{equation}
V(\phi)\simeq V_0+V'_0\phi
\label{taylor}
\end{equation}
where \mbox{$V_0,V'_0=\,$constant} and we have taken the turn-around point to be
at \mbox{$\phi=0$}. Based on the above we can take 
\mbox{$V'/V\simeq V'_0/V_0=\,$constant}, where we have considered that,
because $\phi$ is close to zero, the potential in Eq.~(\ref{taylor})
is dominated by the constant term \mbox{$V\simeq V_0$}. Saturating the 
inequality in Eq.~(\ref{swampland}), we can write the constant logarithmic 
slope as \mbox{$V'/V\simeq c/m_P$}, where \mbox{$c>0$} and without loss of 
generality, we take \mbox{$V'>0$}. Thus, the potential, near the turn-around, 
can be approximated as \mbox{$V=V_0 e^{c\phi/m_P}$}, which in the vicinity of the 
turn-around point, becomes
\begin{equation}
V(\phi)
\simeq V_0\left(1+\frac{c\phi}{m_P}\right)\,.
\end{equation}
The above suggests that \mbox{$V'\simeq cV_0/m_P=$}~constant.

Now, let's suppose that the upslope travelling scalar field dominates the 
Universe. The kinetic density of the field, as it approaches zero, is bound to 
become smaller than the potential density. Thus, we expect inflation to start
(and the field to become the inflaton), 
and we can write the Friedmann equation as \mbox{$V\simeq 3H^2m_P^2$}. 
Now, as we discussed, near the turn-around point, 
\mbox{$V\simeq V_0=\,$constant} and so \mbox{$H\simeq\,$constant}.
The same would be true for the slope of the potential 
\mbox{$V'\simeq cV/m_P\simeq\,$constant}.
Then, the Klein-Gordon equation becomes
\begin{equation}
\ddot\phi+3H\dot\phi+3cH^2m_P=0\,,
\label{KG}
\end{equation}
whose solution is
\begin{equation}
\dot\phi(t)=C\,e^{-3Ht}-cHm_P\,,
\label{solu}
\end{equation}
where $C\equiv\dot\phi(0)+cHm_P$ is an integration constant. We see 
that, only when \mbox{$\dot\phi(0)>0$}, i.e. only when the field is originally 
climbing the upslope of the scalar potential, can we reach a turn-around moment
$t_*$, when \mbox{$\dot\phi(t_*)=0$}. From the above, it is straightforward to 
find that
\begin{equation}
t_*=\frac{1}{3H}\ln\left(\frac{C}{cHm_P}\right)\,.
\label{turnaround}
\end{equation}

As we want to study the system near the turn-around, we Taylor expand the 
expression in Eq.~(\ref{solu}) near $t_*$  and find
\begin{equation}
\dot\phi(t)=\dot\phi(t_*)+\ddot\phi(t_*)(t-t_*)+\cdots\simeq
3cH^2m_P\Delta t\,,
\label{Taylor}
\end{equation}
where we considered that \mbox{$\dot\phi(t_*)=0$}, 
\mbox{$\ddot\phi(t_*)=-3HC e^{-3Ht_*}=-3cH^2m_P$} and we set
\mbox{$\Delta t=t_*-t$}, with the ellipsis denoting higher-order terms in 
$\Delta t$ which we choose to ignore.

To achieve eternal inflation we must satisfy the requirement that
\begin{equation}
|\dot\phi|<\frac{\delta\phi}{\delta t}=\frac{H^2}{2\pi} 
\;\Rightarrow\;\Delta t<\frac{1}{6\pi c m_P}\,.
\label{requir}
\end{equation}
The above shows that the total duration when the system satisfies the condition
for thermal inflation is $2\Delta t$, while approaching the turn-around moment 
and while moving after it. However, in order to have the production of at least
one quantum kick, we need that the total duration, while the system satisfies 
the eternal inflation requirement, is more than a Hubble time, i.e. 
\mbox{$2\Delta t\geq H^{-1}$}.This implies the bound
\begin{equation}
c\leq\frac{H}{3\pi m_P}\ll 1\,,
\end{equation}
which is comparable to the bound in Eq.~(\ref{crange}) and demonstrates that
in order to achieve eternal inflation in this scenario, one still needs a 
shallow slope for the scalar potential. If the slope is large, i.e. 
\mbox{$c\geq 1$}, then the time-interval when the condition of eternal inflation
is satisfied (near the turn-around) is much smaller than a Hubble time so there 
is no time for a single quantum kick.\footnote{This means that the quantum 
displacement of the field is exponentially suppressed.}
Taking into account higher orders in 
$\Delta t$ in the expansion in Eq.~(\ref{Taylor}) only makes matters worse 
because enlarging $\dot\phi$ makes eternal inflation harder to achieve. 
Therefore, it seems that the realisation of the string multiverse through 
eternal inflation is in contradiction with the swampland conjecture in 
Eq.~(\ref{swampland}).

In summary, we have investigated whether steep eternal inflation is possible
and complies with the swampland criterion \mbox{$|V'/V|>c/m_P $} with 
\mbox{$c={\cal O}(1)$}. To achieve steep eternal inflation we considered a 
scalar field, initially climbing up its scalar potential until a turnaround 
point, when momentarily its classical kinetic density becomes zero. We have 
found that the time period when the quantum variation of the field becomes
dominant compared to its classical roll is more than a Hubble time only when
\mbox{$c\ll 1$} and the slope is small enough for the turnaround point to lie 
in a diffusion zone. This means that steep eternal inflation is not possible 
and that eternal inflation, under the swampland conjecture, cannot populate the 
string landscape and realise the string multiverse.

\paragraph{Acknowledgements}\leavevmode\\
%
My research
is supported (in part) by the Lancaster-Manchester-Sheffield Consortium for 
Fundamental Physics under STFC grant: ST/L000520/1. I~would like to thank 
C.~Owen for igniting my interest in this topic.


\begin{thebibliography}{10}

\bibitem{eternal}
  A.~D.~Linde,
Mod.\ Phys.\ Lett.\ A {\bf 1} (1986) 81.

\bibitem{eternallit}
  A.~D.~Linde,
Phys.\ Lett.\ B {\bf 175} (1986) 395;
  A.~S.~Goncharov, A.~D.~Linde and V.~F.~Mukhanov,
Int.\ J.\ Mod.\ Phys.\ A {\bf 2} (1987) 561;
  A.~H.~Guth,
J.\ Phys.\ A {\bf 40} (2007) 6811.

\bibitem{landscape}
  L.~Susskind,
In *Carr, Bernard (ed.): Universe or multiverse?* 247-266
[hep-th/0302219].

\bibitem{landscape+}
  N.~Arkani-Hamed, L.~Motl, A.~Nicolis and C.~Vafa,
JHEP {\bf 0706} (2007) 060;
  H.~Ooguri and C.~Vafa,
Nucl.\ Phys.\ B {\bf 766} (2007) 21.

\bibitem{swampconj}
  G.~Obied, H.~Ooguri, L.~Spodyneiko and C.~Vafa,
  arXiv:1806.08362 [hep-th].

\bibitem{swampcosmo}
  P.~Agrawal, G.~Obied, P.~J.~Steinhardt and C.~Vafa,
  Phys.\ Lett.\ B {\bf 784} (2018) 271;
  A.~Achúcarro and G.~A.~Palma,
arXiv:1807.04390 [hep-th];
  S.~K.~Garg and C.~Krishnan,
arXiv:1807.05193 [hep-th];
  A.~Kehagias and A.~Riotto,
arXiv:1807.05445 [hep-th];
  M.~Dias, J.~Frazer, A.~Retolaza and A.~Westphal,
arXiv:1807.06579 [hep-th];
  E.~Ó.~Colgáin, M.~H.~P.~M.~Van Putten and H.~Yavartanoo,
arXiv:1807.07451 [hep-th];
  L.~Heisenberg, M.~Bartelmann, R.~Brandenberger and A.~Refregier,
arXiv:1808.02877 [astro-ph.CO];
  W.~H.~Kinney, S.~Vagnozzi and L.~Visinelli,
arXiv:1808.06424 [astro-ph.CO];
  L.~Heisenberg, M.~Bartelmann, R.~Brandenberger and A.~Refregier,
arXiv:1809.00154 [astro-ph.CO];
C.~Damian and O.~Loaiza-Brito,
arXiv:1808.03397 [hep-th];
  M.~C.~D.~Marsh,
arXiv:1809.00726 [hep-th];
  S.~Brahma and M.~W.~Hossain,
arXiv:1809.01277 [hep-th];
I.~Ben-Dayan,
  arXiv:1808.01615 [hep-th];
  S.~Das,
arXiv:1809.03962 [hep-th];
  D.~Wang,
arXiv:1809.04854 [astro-ph.CO];
G.~D'Amico, N.~Kaloper and A.~Lawrence,
arXiv:1809.05109 [hep-th];
  H.~Matsui, F.~Takahashi and M.~Yamada,
arXiv:1809.07286 [astro-ph.CO];
M.~Motaharfar, V.~Kamali and R.~O.~Ramos,
  arXiv:1810.02816 [astro-ph.CO];
S.~J.~Wang,
  arXiv:1810.06445 [hep-th].

\bibitem{linde}
  Y.~Akrami, R.~Kallosh, A.~Linde and V.~Vardanyan,
arXiv:1808.09440 [hep-th].

\bibitem{swampcrit}
  D.~Andriot,
arXiv:1806.10999 [hep-th];
  C.~Roupec and T.~Wrase,
arXiv:1807.09538 [hep-th];
  D.~Andriot,
arXiv:1807.09698 [hep-th];
  J.~P.~Conlon,
arXiv:1808.05040 [hep-th];
  H.~Murayama, M.~Yamazaki and T.~T.~Yanagida,
arXiv:1809.00478 [hep-th].

\bibitem{Higgsswamp}
C.~Han, S.~Pi and M.~Sasaki,
arXiv:1809.05507 [hep-ph];
  F.~Denef, A.~Hebecker and T.~Wrase,
Phys.\ Rev.\ D {\bf 98} (2018) no.8,  086004;
  K.~Hamaguchi, M.~Ibe and T.~Moroi,
arXiv:1810.02095 [hep-th].

\bibitem{swampeternal}
  H.~Matsui and F.~Takahashi,
  arXiv:1807.11938 [hep-th].

\bibitem{upclimb}
  E.~Dudas, N.~Kitazawa and A.~Sagnotti,
  Phys.\ Lett.\ B {\bf 694} (2011) 80






\end{thebibliography}
\end{document}